# Resistance switching in oxides with inhomogeneous conductivity[*]

Shang Da-Shang (尚大山),[a)b)†] Sun Ji-Rong (孙继荣),[a)] Shen Bao-Gen (沈保根),[a)] and Wuttig Matthias[b)]

[a)] *Beijing National Laboratory for Condensed Matter Physics and Institute of Physics, Chinese Academy of Sciences, Beijing 100190, People's Republic of China*
[b)] *I. Physikalisches Institut (IA), RWTH Aachen University, 52056 Aachen, Germany*

**Abstract:** Electric-field-induced resistance switching (RS) phenomena have been studied for over 60 years in metal/dielectrics/metal structures. In these experiments a wide range of dielectrics have been studied including binary transition metal oxides, perovskite oxides, chalcogenides, carbon- and silicon-based materials, as well as organic materials. RS phenomena can be used to store information and offer an attractive performance, which encompasses fast switching speeds, high scalability, and the desirable compatibility with Si-based complementary-metal-oxide-semiconductor fabrication. This is promising for nonvolatile memory technology, i.e. resistance random access memory (RRAM). However, a comprehensive understanding of the underlying mechanism is still lacking. This impedes a faster product development as well as an accurate assessment of the device performance potential. Generally speaking, RS occurs not in the entire dielectric but only a small, confined region, which results from the local variation of conductivity in dielectrics. In this review, we focus on the RS in oxides with such an inhomogeneous conductivity. According to the origin of the conductivity inhomogeneity, the RS phenomena and their working mechanism are reviewed by dividing them into two aspects: interface RS, based on the change of contact resistance at metal/oxide interface due to the change of Schottky barrier and interface chemical layer, and bulk RS, realized by the formation, connection, and disconnection of conductive channels in the oxides. Finally the current challenges of RS investigation and the potential improvement of the RS performance for the nonvolatile memories are discussed.

---

[*] Project supported by the National Nature Science Foundation of China, the National Basic Research of China, the Knowledge Innovation Project of the CAS, and the Alexander von Humboldt Foundation (for S.D.S).
[†] Corresponding author. E-mail: shangdashan@iphy.ac.cn

# 1. Introduction

The electric-field-induced resistance switching (RS) effect describes a situation where the resistance of a dielectric material or devices consisting of a metal-insulator-metal (MIM) sandwich configuration can be switched repeatedly between high and low resistance states by applying an electric field. The resulting resistance states can be maintained for a long time after removing the external field. Different from the traditional dielectric breakdown phenomena, which lead to a permanent resistance decrease so that switching back is impossible, the RS process is reversible and can be repeated a certain number of times. Depending on the external bias polarity, the RS can be induced either by unipolar or by bipolar forms, as shown in Fig. 1. For the former, the switching direction depends on the polarity of the applied electric field, while for the latter there is no polarity dependence but only a dependence on the amplitude of the applied voltage.

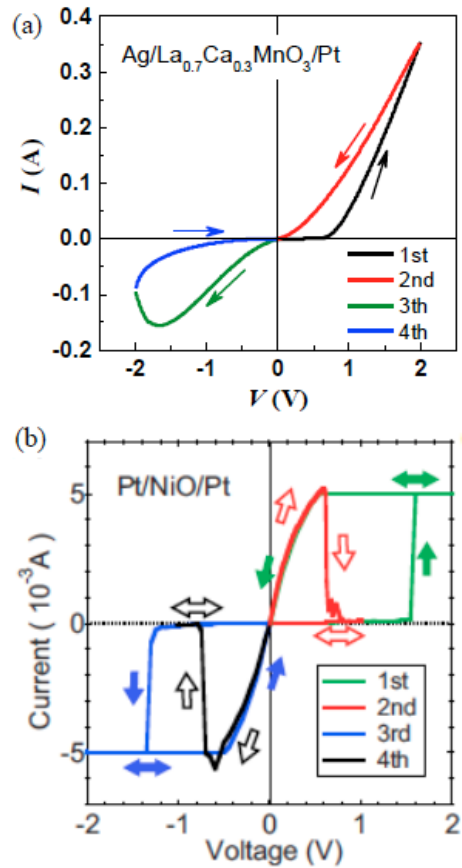

Fig. 1 $I$-$V$ curves for (a) bipolar RS in a Ag/La$_{0.7}$Ca$_{0.3}$MnO$_3$/Pt cell,[30] and (b) unipolar RS in a Pt/NiO/Pt cell.[15] Copyright 2006 and 2008 The American Physical Society.

The RS phenomenon can be traced back more than 50 years ago. The first papers on RS in oxide insulators, were published by Kreynina,[1] and Hickmott.[2] Then a first peak in the number of investigations occurred until the 1980s, when almost all the transition metal oxides where studied in an MIM configuration arose. Dearnaley *et al*,[3] Oxley,[4] and Pagnia,[5] comprehensively reviewed the research results obtained in this period and proposed several models to explain the RS effect. These activity faded because of the lack of progress in understanding and controlling the phenomena due to insufficient analytical tools and technological equipment at that time. The current renewed interest in this area started since around the year of 2000. Prompted mainly by the experimental results of Kozicki *et al* on ion-doped chalcogenides,[6] Liu *et al* on manganites,[7] and Beck *et al* on perovskites,[8] a new application for nonvolatile memory called resistance random access memory (RRAM) has been introduced, which is promising for its easy fabrication, fast switching speed (<100 ns), high scalability, and good compatibility with the Si-based complementary-metal-oxide-semiconductor (CMOS) technology. Furthermore, since it is possible to switch the device to several resistance states, RRAM has the potential to realize a multilevel memory,[8,9] in which the storage density of one memory cell will multiply without changing cell volume. In 2008, Strukov *et al* at Hewlett-Packard labs have successfully identified the resistance switching phenomena as memristive systems,[10] and subsequently proposed in a broader range of applications including the Boolean logic implementations and neuromorphic systems.[11,12] Driven by these opportunities, a large variety of materials have been demonstrated to have the RS property and have the possibility to be the candidates for RS-related device fabrication, including perovskite oxides,[13,14] binary transition metal oxides,[15,16] wide band gap high-k dielectric oxides,[17,18] higher chalcogenides,[19,20] and carbon-based materials.[21,22]

The major roadblock to commercialize RS-related devices is the lack of knowledge regarding the RS physical mechanism, even though models, as reviewed in detail by Waser *et al* and Sawa,[23-25] have been proposed based on different type of material properties. From the viewpoint of underlying physics, these models could be classified into two categories. One is based on an electronic mode such as a Mott metal-insulator transition,[26] ferroelectric polarization,[27,28] carrier trapping/detrapping,[29,30] and the other one relies on an ionic model, such as cation- and anion-migration,[31,32] ion valence change,[33] interfacial Schottky barrier narrowing,[34] random breakdown,[35] and metastable phase generation.[36,37]

No matter which mechanism dominates the RS behavior, a noteworthy feature for almost all the RS phenomena in oxides is that the RS occurs not in the whole cell but only in a small, confined region. The presence of this local active region in the RS cells implies that the conductivity distribution in the cell has to be inhomogeneous. Two different types can be distinguished depending on the location of the inhomogeneity. In one case RS takes place at the interface between the metallic electrodes and the dielectric oxides, while in the second case RS occurs inside the bulk of the oxides usually accompanied by the formation and rupture of conductive channels, as shown in Fig. 2. Obviously, the origin, distribution, and manipulation of the inhomogeneous conductivity are not only highly relevant for the understanding of the microscopic RS mechanism but also for the improvement and estimation of the performance potential for future non-volatile memory devices. Therefore, the main focus of this review is on the inhomogeneous conductivity in the RS systems and its relation with the RS mechanism. We will discuss this topic for the two different classes, interface RS and bulk RS. Firstly, the interface RS will be presented, focusing on the roles played by the Schottky-like barrier and the interface layer, and subsequently the interface inhomogeneity. Secondly, the intrinsic defect and electroforming as the origin of the inhomogeneous conductivity inside the bulk of oxides will be presented. Then the bulk RS behavior will be discussed by means of observation of conductive channels, which are typical for the inhomogeneous conductivity in oxides. The review will be conclude with the current challenges of RS investigation and the outlook of the potential performance of the RS phenomena in the application of nonvolatile memories.

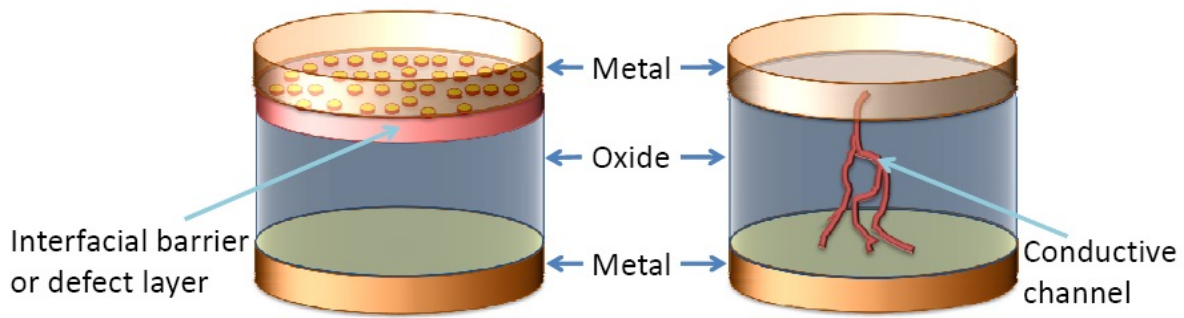

Fig. 2 The inhomogeneous RS behavior can be classified depending on the location of the inhomogeneous region. (left side) Interface-type, in which RS occurs in the metal/oxide interface region; (right side) Bulk-type, in which RS occurs in oxides in form of conductive channels.

## 2. Resistance switching at metal/oxide interfaces

Usually the interface RS occurs in a metal/oxide/metal sandwich cell, where the oxides originally have a relatively high bulk conductivity. Hence, prior to talk about the interface RS, it is first necessary to clarify the origin of the contact resistance, which can be changed by applying an electric field.

### 2.1 Schottky-barrier-type RS

The origin of the high contact resistance between metal electrode and oxide films can be mainly attributed to two reasons. One reason is due to the formation of a Schottky-like barrier. In this case, the RS is usually ascribed to a field-induced variation of the interfacial barrier. Fig. 3(b) and 3(c) show a current-voltage (*I-V*) characteristics of two Au/Nb:SrTiO$_3$ junctions with 0.05 and 0.5 wt% Nb doping concentrations, respectively.[38] The cells were prepared by depositing gold electrodes on the Nb:SrTiO3 single crystals by magnetron sputtering, and then depositing titanium electrodes on the single crystals by pulsed laser deposition. From the semi-logarithmic *I-V* characteristics, it can be seen that the resistance of the Ti-Ti structure is very low, and the *I-V* curves show an Ohmic-like transport behavior without any hysteresis, meaning that there is no Schottky barrier at the Ti/Nb:SrTiO$_3$ interface. However, once one Ti electrode was change by gold, the resistance increased largely and an obvious *I-V* hysteresis is observed. The positive branch of the log*I-V* curve shows a linear trend and fits well to with the Schottky barrier function $J_F \propto exp(qV/nk_BT)$, where $J_F$ is the forward current density, $q$ is the electron charge, $n$ is ideal factor, $k_B$ is the Boltzmann constant, and $T$ is the absolute temperature. Compared with the Ti-Ti structure, it can be concluded that the large device resistance value comes from the Au/Nb:SrTiO$_3$ interface, where the Schottky-like barrier is formed. The large *I-V* hysteresis demonstrates that nonvolatile RS takes place. This type of RS is bipolarity, that is, the junction resistance increases in the positive bias region and decreases in the negative bias region.

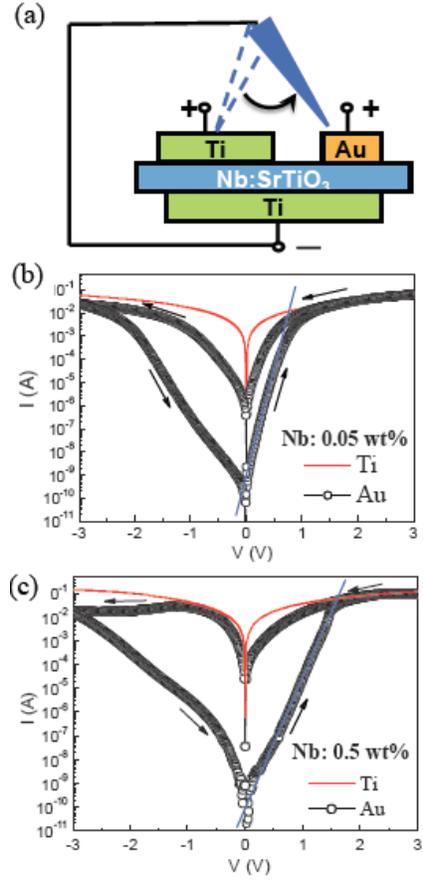

Fig. 3 (a) Schematic for the electrode setting and measurement of metal/Nb:SrTiO$_3$ junctions. Semilogarithmic *I-V* characteristics of metal/Nb:SrTiO$_3$ junctions with (a) Nb: 0.05 wt% and (b) Nb: 0.5 wt% for the voltage cycling of 0→3 →0 →-3 →0 V. Arrows indicate the sweeping direction.[38]

The RS behavior can also realized by applying electric pulses. As shown in Fig.4, a multilevel RS is obtained in the Au/Nb(0.5 wt%):SrTiO$_3$ junctions by changing the polarity and amplitude of electric pulses (pulse width=1 ms). The junction resistance shows a high to low switching when induced by positive pulses, whereas a reverse switching occurs upon reversing the electric polarity. The RS between different resistance states is a transient process followed by a slow relaxation of junction resistance. However, each obtained resistance state cannot restore to its last state and will become stable after the relaxation (usually an hour). By applying a sequence of voltage pulses with appropriate amplitudes, a stable and reproducible multilevel RS can be obtained. This multilevel RS can be used to realize a multilevel memory. For example, negative-pulse-induced LRS to HRS is appropriate for a SET process in which several resistance states can be obtained to save information of "1", "2", "3", "4", while positive-pulse-induced HRS to LRS is appropriate for a RESET process, in which the resistance can be restored to the initial state to save the information "0".

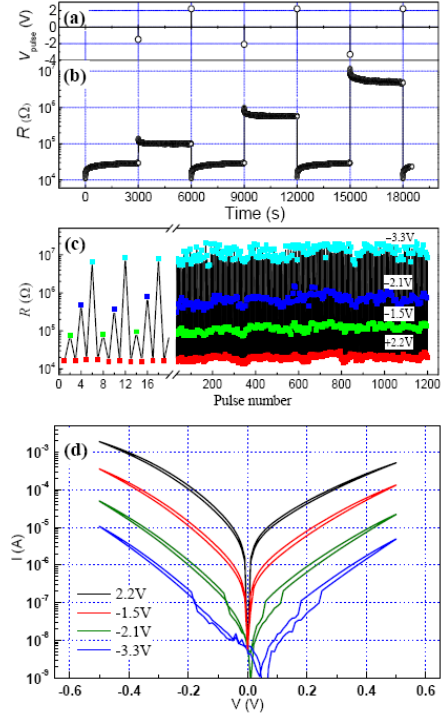

Fig. 4 Multilevel RS of the Au-Nb(0.5 wt%):SrTiO$_3$ junctions with a sequence of voltage pulses of given polarities. (a) The sequence of voltage pulses with voltage varied as +2.2, -1.5, -2.1, and -3.3 V. Pulse width is fixed at 1 ms. (b) Response of the junction resistance to the sequence of voltage pulses shown in (a). Readout voltage is 0.1 V and readout time is 3000 s. (c) Multilevel RS reproducibility of Au-NSTO junctions. (d) *I-V* curves of different resistance states obtained by electric pulsing.[38]

Many mechanisms have been reported to be possible to cause the Schottky barrier change and subsequently the RS behavior. They include polarity changes in ferroelectric oxides,[27,39] charge trapping or releasing of the defects in the depletion layer,[40] resonant tunneling through the barrier,[34] the presence of a low dielectric layer and interface states,[41] and field-induced drift of dopants.[25,42] Among them, the field-induced oxygen vacancy migration model seems to be most widely accepted and much convincing experimental evidence has been given. As shown in Fig. 5(a), Sawa *et al* determined the potential profiles of the depletion layer in Ti/Pr$_{0.7}$Ca$_{0.3}$MnO$_3$ Schottky junctions by capacitance-voltage (*C-V*) measurements.[25] The *C-V* curves show a hysteretic behavior and *C* is larger in the LRS than in the HRS. This suggests that the depletion layer width is narrower in the LRS than in the HRS. Thus, electrons possibly pass through the thin barrier via a tunneling process in the LRS, whereas the thick barrier in the HRS prevents tunneling. The drift of oxygen vacancies can explain the change of the barrier width. This model has also been successfully used to explain the RS behavior in the SrRuO$_3$/Nb:SrTiO$_3$ junctions, which shows an opposite RS polarity to the Ti/Pr$_{0.7}$Ca$_{0.3}$MnO$_3$ junction. Yang *et al* also adopted the influence of oxygen vacancy drift on the barrier width to interpret the RS behavior in their Pt/TiO$_2$/TiO$_{2-x}$/Pt systems.[42,43] As shown in Fig. 5(b), under an applied electric field, oxygen vacancies can drift into the interface region, reducing the electronic barrier and then resulting in a LRS. Under an electric field with the opposite polarity, the oxygen vacancies are repelled away from the interface region, recovering the electronic barrier to regain the HRS. Yang's model can be also used to explain other binary and perovskite RS systems.[44,45] Although both Sawa's model and Yang's model are based on the oxygen vacancy migration, the RS polarities according to the experimental results explained by the two models are opposite each other. For example, the forward bias of the Schottky-like barrier corresponds to the resistance decreasing in Sawa's model, while it corresponds to the resistance increasing in Yang's model.

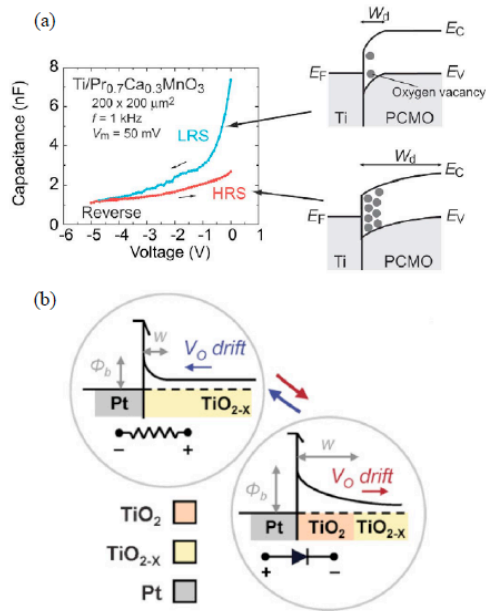

Fig. 5 (a) Capacitance-voltage curves under reverse bias for a Ti/Pr$_{0.7}$Ca$_{0.3}$MnO$_3$/SrRuO$_3$ cell and corresponding change of the depletion layer width ($W_d$) at the Ti/Pr$_{0.7}$Ca$_{0.3}$MnO$_3$ interface by applying an electric field.[25] Copy right 2008 Elsevier. (b) Schematics of the modes of RS at the Pt/TiO$_2$ interface.[43] Positive charged oxygen vacancies drift towards or away from an interface depending on the applied voltage polarity, resulting in RS between Ohmic and Schottky-like contact. Copyright 2009 Wiley-VCH.

The electronic transport behaviors has been investigated for the RS mechanism of the Au/Nb:SrTiO$_3$ junctions.[46] There are mainly two different currents through a Schottky junction. The first one arises from electron tunneling and the second one from thermionic emission. To clarify the relevance of these two transport channels, the resulting *I-V* characteristics were investigated under different temperature conditions. Three types of samples were selected. The first one has an Nb doping level of 0.5 wt%, the second one has an Nb doping level of 0.05 wt%, and the third one has an Nb doping lever of 0.05 wt% but utilizes a single crystal, which was annealed in oxygen atmosphere before the measurement. As shown in Fig. 6, the log*I-V* slope of both the first sample and the second samples show a temperature independent behavior. This feature is consistent with the predictions based on the Newman equation and strongly suggests the electron tunneling behavior. In contrast, the log*I-V* slope of the third sample shows a pronounced temperature dependence, as expected for thermal emission behavior. Referring to the *I-V* characteristics, it is interesting to find that the RS behavior occurs only in the junctions with the electron tunneling transport.

Occurrence of electron tunneling implys that the depletion layer width is quite thin and a considerable number of defects that assist tunneling must exist in this layer. When an external bias is applied, a thinner depletion width leads to a higher effective electric field in the depletion layer, which may drive the migration of the charged defects in the depletion layer to form conductive channels. Since the post-annealing in oxygen atmosphere will eliminate the donor-like oxygen vacancy defects in the vicinity of the interface, the depletion layer of the annealed-Au/Nb:SrTiO$_3$ junction becomes wider. Meanwhile, the charged defect density in the depletion layer was also reduced. In this case, it becomes difficult to drive the defects to form the conductive channel, until the permanent breakdown takes place. Therefore, the RS phenomena disappeared in the post-annealed sample. This scenario is similar to Yang's model if considering the dopants as oxygen vacancies.[42] However, as has been said above, applying Yang's model to explain the RS of the Au/Nb:SrTiO$_3$ Schottky junction would lead to the opposite polarity. That means some other mechanism, as will be discussed later, might exist in the interface RS based on a Schottky-like barrier.

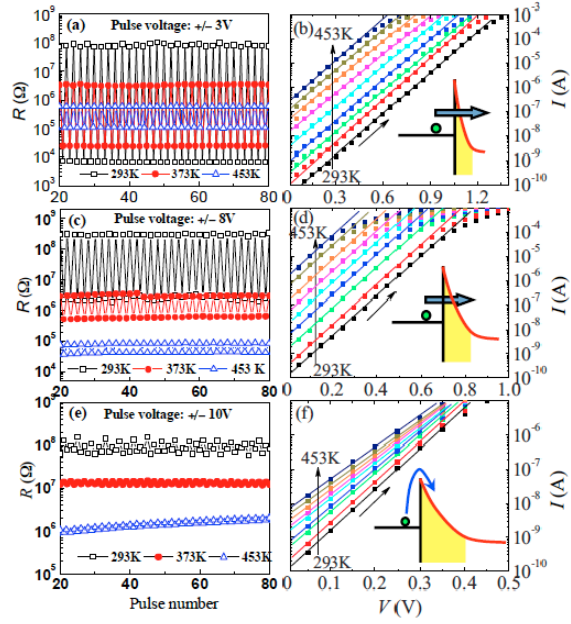

Fig. 6 RS and *I-V* characteristics of [(a) and (b)] Au/Nb(0.5 wt%):SrTiO$_3$ junction, [(c) and (d)] Au/Nb(0.05 wt%):SrTiO$_3$ junction, and [(e) and (f)] Au/Nb(0.05 wt%):SrTiO$_3$ junction annealed in oxygen atmosphere at 823 K for 30 min.[46] Copyright 2009 American Institute of Physics. Inset shows the schematics for the different electron transport modes.

## 2.2 Interfacial-layer-type RS

The other reason to form a high contact resistance is due to the interfacial layer formation caused by the chemical reaction between metal and oxide. Ideally, the Schottky barrier height is defined as the difference between the electron affinity of the semiconductor and work function of the metal and is supposed to be proportional to the work function of the metal. However, in reality there may be an interfacial layer between the metal and semiconductor, which changes the band structure matching at the interface and also causes a voltage drop through the layer, leading to a deviation the Schottky barrier height from the ideal value. Yang *et al* identified the role of the interfacial layer in the current injection through the metal/oxide interface for TiO$_2$ in contact with various electrode metals (Au, Pt, Ag, Ni, W and Ti).[47] As, shown in Fig. 7, no clear relation can be seen between the resistance of the devices and the work function values of the top electrode metals. According to the Ellingham diagram, which shows the Gibbs free energy formation of metal oxides and can be used to predict the possibility that reactions may occur, Ni, W, and even Pt electrodes can reduce the TiO$_2$ to a Magnéli phase of Ti$_n$O$_{2n-1}$, and induce oxygen vacancies in TiO$_2$ films by the formation of NiO, WO$_3$, and PtO$_2$. Fig. 7(c) shows the resistance of the cells plotted vs. the standard free energy of formation of the electrode metal oxides, where an approximately monotonic trend can be seen, indicating that the larger the formation free energy of the metal electrode oxide, the smaller the resistance of the devices with that metal electrode.

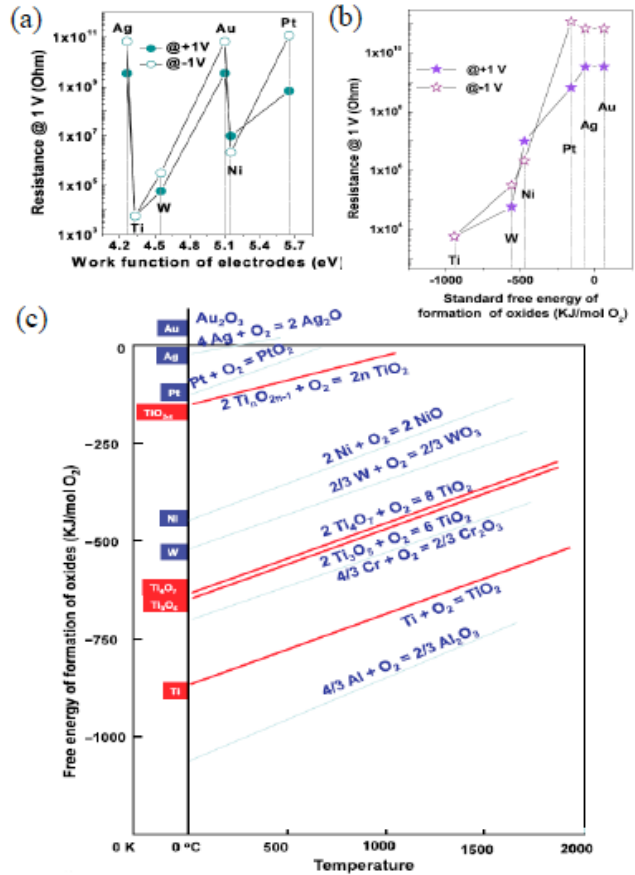

Fig. 7 (a) Resistance of metal/TiO$_2$/Pt cells at a read-out voltage of +/- 1 V vs. (a) the work function and (b) the standard free energy of oxide formation of the top electrode metals. (c) Ellingham diagram with different TiO$_2$ formation reactions from its sub-oxides.[47] Copyright 2011 Springer.

This tendency was also found by Liao et al[48] and Yamamoto et al,[49] in other perovskite oxide systems, where the interfacial layer was observed directly by high-resolution transmission electron microscopy (HRTEM) and photo emission spectroscopy (PES) measurements in the films with active electrodes. However, the RS behavior in these films is dominated not by the Schottky barrier change but by the concentration change of oxygen vacancy concentration in the interfacial layer. Fox example, a positive voltage applied to the Ti electrode will pull out the oxygen vacancies in the $TiO_x$ layer and oxidation of $TiO_x$ region causes the switching to the HRS. When a negative bias is applied to the Ti electrode, oxygen vacancies will be attracted back to the $TiO_x$ layer and the reduction of the $TiO_x$ causes a switching back to the LRS. This model can also be used for the RS systems with other metal electrodes such as Al, W, Ta, where $AlO_x$, $WO_x$ and $TaO_x$ form at the metal/film interface region.[50-53]

The interfacial layer formation is not only related to the Gibbs free energy of oxide formation of metal electrodes, but also with the electrode preparation techniques. For instance, Shang et al[54] and Fujimoto et al[55] used silver paste as the top electrode and found that the Ag(paste)/$La_{0.7}Ca_{0.3}MnO_3$ interface exhibit an obvious *I-V* hysteresis and a stable RS property. In contrast, Sawa et al[56] and Liao et al[48] found no RS phenomenon in the Ag/$Pr_{0.7}Ca_{0.3}MnO_3$ interface, in which an Ag top electrode was prepared by electron beam and magnetron sputtering. Yang et al found the RS endurance in the Ag/$La_{0.7}Ca_{0.3}MnO_3$/Pt cell can be improved by alloying Al into the Ag electrode during electrobeam evaporation.[57] These contradictions might originate from the different chemical states of silver at the silver/oxide interface. For example, Ag is commonly used as an inert electrode material due to its antioxidation property. However, Ag can be readily oxidized during Ag film deposition by some physical methods, if there is oxygen available in the growth chamber. To clarify the effect of an Ag oxide on the RS behavior, three types of devices were prepared with the structure of Ag/$Ag_2O$/Pt (AOP), Ag/$WO_{3-x}$/Pt (AWP), and Ag/$Ag_2O$/$WO_{3-x}$/Pt (AOWP), respectively.[58] As shown in Fig. 8, firstly the original resistance of AOWP is much higher than that of AOP and AWP, indicating that the interface of $Ag_2O$/$WO_x$ generates a high contact resistance. The rectifying *I-V* curve and linear semilogarithmic current at the positive bias of AOWP indicate that an Schottky-like barrier exists at the $Ag_2O$/$WO_x$ interface. *I-V* hysteresis only appeared in the AOP and AOWP devices, meaning that the $Ag_2O$ layer plays a dominant role in the RS. From the micro-Auger spectra of the sample with HRS and LRS, it can be seen that the $AgO_x$ layer actually is inhomogeneous and composed of a mixed phase containing Ag and $Ag_2O$. The RS can be ascribed to the content decreasing (increasing) of the $Ag_2O$ (Ag) in the mixed layer due to the reversible electrochemical redox reaction of $Ag(Ag_2O) + e^- \leftrightarrow Ag$ under the electric field, which changes the contact barrier at the $Ag_2O$/$WO_x$ interface.

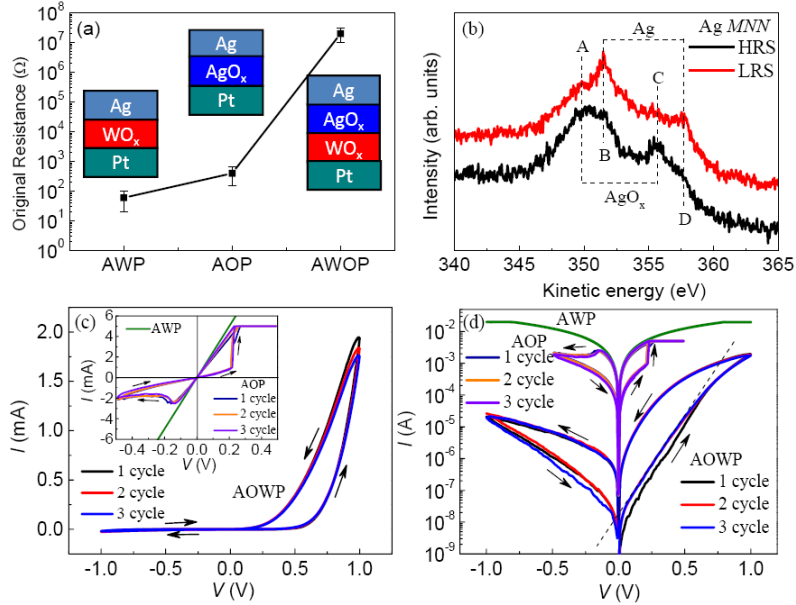

Fig. 8 (a) The original resistances of the Ag/WO$_x$/Pt, Ag/AgO$_x$/Pt, and Ag/AgO$_x$/WO$_x$/Pt cells. (b) Ag MNN Auger spectra of the Ag/AgO$_x$/WO$_x$/Pt cell under HRS and LRS. The HRS and LRS were obtained by applying an electric field to a removable Ag probe with a diameter of ~20 μm, which contacted directly with the AgO$_x$ surface and was removed after each RS. (c) and (d) I-V characteristics in decimal and semilogarithmic patterns.[58] Copyright 2011, American Institute of Physics.

**2.3 Is interface RS homogeneous or inhomogeneous?**

In the search for promising oxide materials for future nonvolatile memories, special attention has to be paid to their scaling capability. The issue of scaling is strongly linked to the question whether the RS is distributed homogeneously over the cell area or localized to one or a few conducting channel parts. Generally, homogeneous RS devices show a clear scaling of the change of resistance with the electrode area. Interface-layer-type RS systems usually shows the homogeneous RS, as reported by Meyer *et al*,[59] Hasan *et al*,[60] and Sawa *et al*.[25,61] When the interface-layer-type RS shows some inhomogeneous RS behavior, this resembles the conductive channel behavior in the oxide bulk, which will be discuss later in Section 3. Sim *et al* reported on a so called homogeneous barrier-type RS in the Pt/Nb:SrTiO$_3$ Schottky junctions.[62] The resistance values in On and Off states depend linearly on the electrode area, suggesting that the RS takes place over the entire area of the interface. In contrast, Yang *et al* suggested that RS involves local changes to the Schottky barrier at the Pt/TiO$_2$ interface.[42,63] They cut the active electrode into several parts after forming and RS, and found only one of the parts shows the RS phenomenon while other parts behave approximately virgin. This clearly demonstrates that after electroforming typically one single conductive channel exists. Under an applied electric field, the positively charged oxygen vacancies drift through the most favorable diffusion paths, such as grain boundaries, to form channels with a high electrical conductivity. Once one or more conducting channels penetrate the electronic barrier, the device is switched to LRS. Applying a voltage with the reverse polarity repels the vacancies in the conducting channel away from the top interface and the original electronic barrier is recovered. Szot *et al*[45] adopted a metallic tip with the diameter of about several nanometer as one of the electrodes to form a point contact system on a SrTiO$_{3-\delta}$ single crystal and demonstrated the RS of the interfacial state between a Schottky barrier and a metallic contact upon applying electric pulses with different polarity, as shown in Fig. 9. The experiments gave clear evidences that the conductivity of the SrTiO$_3$ single crystal surface is inhomogeneous due to the presence of a high density of dislocations, which provide fast channels for oxygen migration. RS occurs in these channels and thus only needs to a very small region (~2nm in diameter) of the oxide surface.

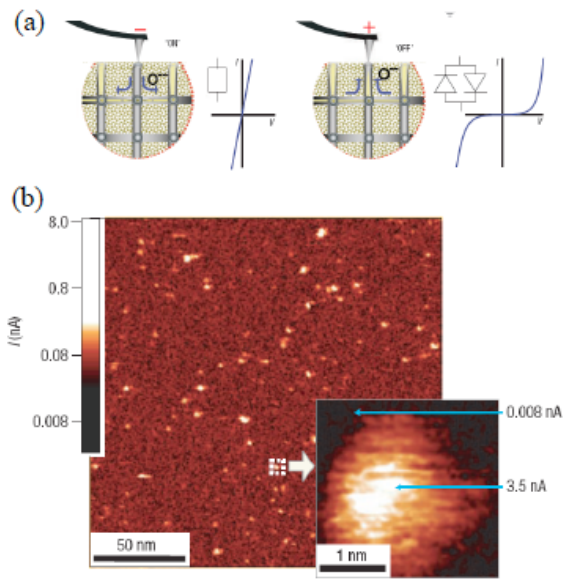

Fig. 9 Schematic illustration of the electromigration of oxygen in the upper segment of a SrTiO$_{3-x}$ single crystal upon the RS. The electrical properties can be characterized by a resistor with linear *I-V* characteristics and a diode with symmetrical nonlinear *I-V* characteristics. (b) Conductivity map of the surface of a SrTiO$_{3-x}$ single crystal as recorded by C-AFM. Conductive channels are present on the surface. Inset shows the conductive channel with a dimension of 1-2 nm, corresponding to the size of the core of a typical edge-type dislocation.[45] Copyright 2006 Nature Publishing Group.

The photovoltaic effect is a distinctive property of a Schottky junction. The electric signal is generated by the built-in field of the Schottky barrier. Hence, no external electric-field is needed to detect the barrier change in this case. Fig. 10 shows the open-circuit photovoltage ($V_{noc}$) and short-circuit photocurrent ($I_L$) of the Au/Nb(0.5 wt%):SrTiO$_3$ junction illuminated by a laser with a wavelength of 460 nm. The strong dependence of $V_{noc}$ on the junction resistance $R_j$, which was changed from 900 MΩ to 70 KΩ by applying electric pulses, suggests the variation of the barrier potential.[64] Different from $V_{noc}$, however, $I_L$ exhibits an almost invariant value regardless of the change of $R_j$, indicative not only of the invariance of the barrier height/width but also of the carrier diffusion lengths. Different laser sources with wavelengths between 380 to 940 nm have also been used to check $I_L$. All the results show the independence of $R_j$.[65]

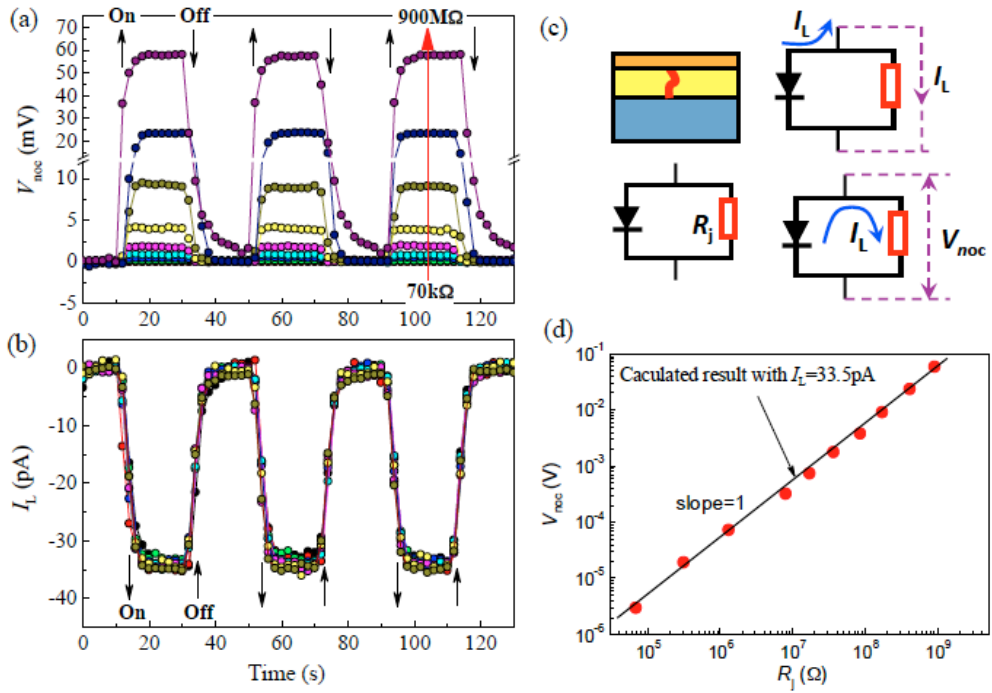

Fig. 10 (a) Open-circuit-photovoltage ($V_{noc}$) and (b) short-circuit-photocurrent ($I_L$) produced by laser (460 nm) illumination for the Au/Nb(0.5%):SrTiO$_3$ junction in different resistance states, which were obtained by applying voltage pulses with an amplitude of 3-6 V and width of 1 ms. (c) Schematic of the equivalent circuit consisting of a diode in parallel with a resistor. (d) $V_{noc}$ as a function of junction resistance ($R_j$).[64] Copyright 2008 American Institute of Physics.

An additional conductive channel through the Schottky barrier is assumed to explain the contradictory behaviors between the obtained $V_{noc}$ and $I_L$. In this case, the equivalent circuit of the junction can be seen as a diode in parallel with a resistor. The resistor represents the conductive channels, and the diode represents the remaining unchanged region of the Schottky barrier. The application of voltage pulses changed the value of the resistor but has little effect on the diode during the resistance switching. As shown in Fig. 10(c), the appearance of the channels has no obvious effect on $I_L$, since the total area of the conductive channels could be much smaller than that of the interface. However, when measuring the $V_{noc}$ the conductive channels will short-circuit the interfacial barrier and carry the main current through the junction. Thus, the measured photovoltage is not a real open circuit signal at all but a measure of the voltage on the conductive channels. The measured photovoltage can be expressed as $V_{noc} = (I_L - I_F)R_j$. Since $I_L \gg I_F$ at low bias, the equation can be simplified as $V_{noc} \approx I_L R_j$. As shown in Fig. 10(d), $V_{noc}$ indeed shows a linear increase with $R_j$ at a rate of $I_L \approx 34$ pA, which is a value consistent with the experimental data of ~33.5 pA in Fig. 10(b). These photoresponse results provide a clear proof that the RS of the Au/Nb:SrTiO$_3$ junction takes place at a local region of the interface. In fact, Yang *et al* also gave the similar diode-$R_j$ parallel model for the Pt/TiO$_2$/TiO$_{2-x}$/Pt Schottky barrier system where they denote the $R_j$ as memristor.[42]

A more complicated interface RS was reported by Dittmann *et al*.[66,67] By using a delamination technique to remove the top electrode and then using a conductive-tip atomic force microscopy (C-AFM), the investigation of the active RS interface with nanoscale lateral resolution is realized. As shown in Fig. 11, two types of RS, filament-RS and area-dependent RS were observed after electroforming in the same device with Pt/Fe:SrTiO$_3$/Nb:SrTiO$_3$ structure. Interesting enough, the two types of RS show opposite RS polarity. This peculiar RS behavior was also observed in other related systems.[68] Both types of RS are interpreted by the oxygen vacancy migration driven by an external electric field, as shown in Fig. 11 (e) and (f). In the case of area-dependent RS, oxygen vacancies will start to migrate out of the lower regions of the film and accumulate at the upper surface, forming a reduced region under negative bias. At the same time, the further supply of vacancies from below is hindered because the electric field strength is only sufficient within the first 10 nm below the surface. An oxygen vacancy depleted region is formed within the film that has very poor conductivity and therefore causes the overall sample resistance to rise to the HRS. Under the positive bias, oxygen vacancies are pushed back into the depleted region. At the same time, new vacancies can be created at the surface due to oxygen release. The overall sample resistance is decreased and the junction switches back into the LRS. The area-dependent RS therefore shows "eightwise" switching polarity. The main difference of the filament-RS mechanism from the area-dependent RS is the higher oxygen mobility in the filament due to the existence of a high density of dislocation defects, which can play a role as a fast ion channel. As shown in Fig. 11 (f), in this case, the depletion region of oxygen vacancies will not form; instead oxygen vacancies would immediately close the gap. The attraction of oxygen vacancies into the upper interface under negative bias corresponds to lowering of the Schottky-like barrier (LRS), and the repulsion of oxygen vacancies under positive bias corresponds to recovery of that barrier (HRS). The filament therefore switches with the "counter eightwise" polarity. This model seems to give a comprehensive explanation for different RS systems with different polarity. Meanwhile, it also indicates that the Schottky-barrier-type RS should be more complicated and inhomogeneous than what has been reported and the redox reaction as discussed in interfacial-layer-type RS also needs to be reconsidered.

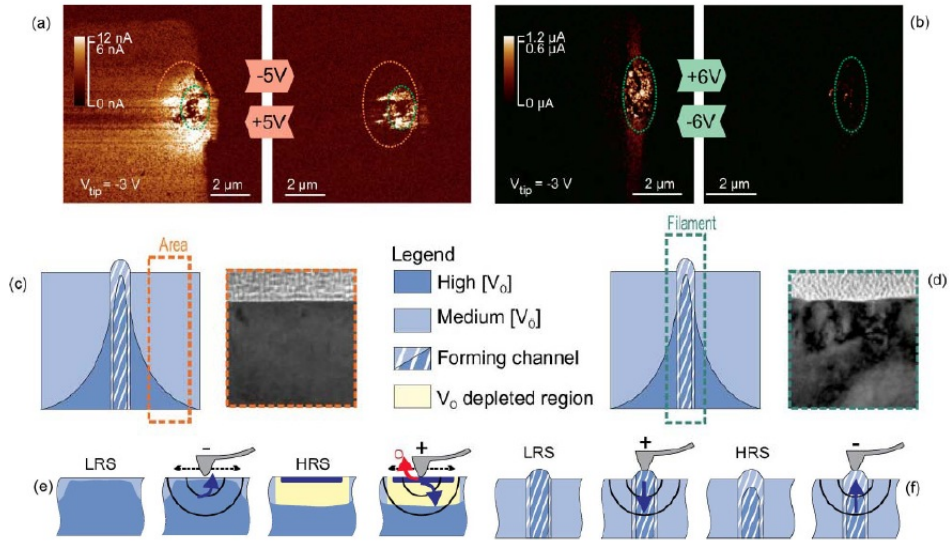

Fig. 11 (a) Homogeneous-type and (b) filament-type tip-induced RS of a delaminated Fe(1 at%): SrTiO$_3$/Nb(1 at%):SrTiO3 junction. (c) Schematic of the sample crosses section with a marker for the homogeneous RS region and corresponding TEM image. (d) Schematic of the sample cross section with a marker for the filament RS region and corresponding TEM image, which shows a strongly increased defect density. (e) and (f) shows the schematic illustration for the homogeneous and filament AFM-based RS mechanism, respectively.[66] Copyright 2012 IEEE.

## 3. Resistance switching inside oxides (conductive channels)

RS inside oxides, called as 'bulk RS', usually takes place in a metal/oxide/metal cell, where the oxides originally have relatively low conductivity and bear the main electric stress under the bias condition. However, it doesn't mean that RS takes place in the whole bulk region. Instead, almost all the reported bulk RS exhibit a localized feature. The active region for RS usually is named as the conductive channel (or filament), which usually only occupies a small volume compared with the bulk, as shown in Fig. 2 (b), and can be considered as an inhomogeneous conductive distribution in the oxides. Moreover, bulk RS doesn't mean it only involves the oxide itself. In some cases, the interface properties, such as Schottky-like barrier and interfacial chemical reaction discussed above also have some influence on the RS behavior since they can change the conductive homogeneity in the cell. Therefore, it is essential to have a good understanding of the origin of the inhomogeneous conductivity in oxides.

### 3.1 Intrinsic inhomogeneous conductivity in oxides

As reported by Szot *et al*,[45] (see Fig. 9) the inhomogeneous conductivity in single crystals can be caused by dislocations, which have been demonstrated to possess a higher conductivity compared to that of the surrounding matrix. *Ab initio* calculations indicate that the high conductivity of dislocations is due to the reduction of oxygen occupied in the dislocation core, leading to a local change of the Ti valence.

Besides single crystals, many materials exhibit RS behaviors in the polycrystalline structure, in which grains are packed together, separated by grain boundaries. Grain boundaries can also be considered as defects and should have an important effect on the local conductivity and RS behavior. Fig. 12 shows topography and corresponding current images of the polycrystalline $WO_{3-x}$ films on F-doped tin oxide substrates, which are annealed at 400 °C in oxygen atmosphere for 30 min before measurement.[69] The current images are measured by applying external bias on the CAFM tip, which serves as one of the electrodes to examine the local conductivity. It can be seen clearly that the regions corresponding to the grain surface exhibit a higher conductivity than that of the regions corresponding to the grain boundary surfaces. *I-V* characteristics on the grain and grain boundary surfaces indicate that RS occur only in the grain region. The current mapping of HRS and LRS further confirmed reversible RS in the grain region with the diameter size of ~10 nm, as shown in the inset of Fig. 12 (d).

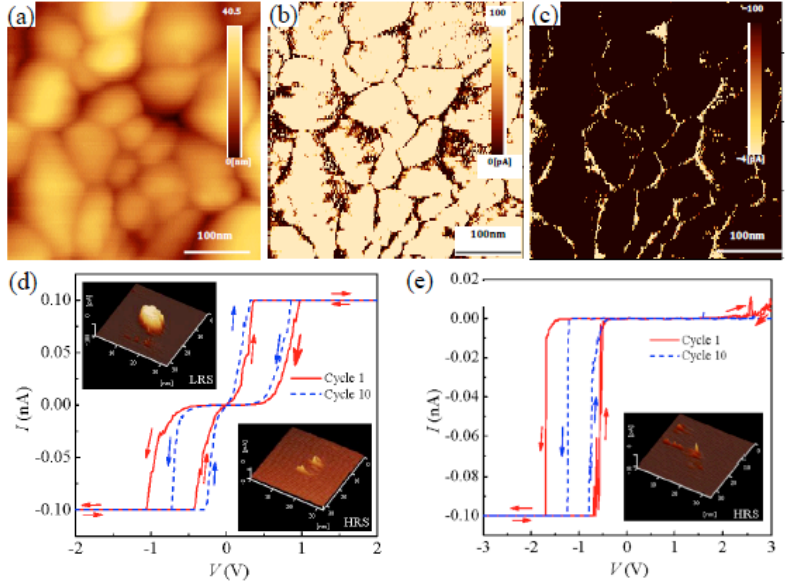

Fig. 12 (a) Topography of the $WO_{3-x}$ film on F-doped tin oxide/glass substrate. (b) and (c) show the corresponding current images with tip bias of 2 V and -1.5 V. (d) and (e) show the local *I-V* curves for grain and grain boundary regions, respectively, measured by the CAFM tip with external biases ranging from -3 to 3 V. Upper left and lower right insets in (a) show the current mappings after negative and positive voltage sweeps, respectively. Inset in (b) shows the current mapping after negative voltage sweep. The current mappings were obtained for a tip bias of -0.5 V.[69] Copyright 2011 IOP Publishing Ltd.

By scanning the CAFM tip with an external bias larger than the RS threshold voltage, an entire area can be selected to switch reversibly between the HRS and the LRS, as shown in Fig. 13. A two-fold overwriting process was performed. First, a 0.6×0.6 μm² area in the center of a 1×1 μm² area was scanned under a tip bias of -3 V. Then a smaller inner area 0.3×0.3 μm² with the same center as the previous image was scanned under a tip bias of 3 V. The bright contrast indicates the LRS within the square region and the HRS outside. Some dark fragments corresponding to the grain boundaries in the LRS region are also observed due to the non-switching property of the grain boundaries. Besides grain boundary regions, it should be noted that no full area scanned by the biased-tip show RS property. It has been demonstrated that the RS area increases with increasing the tip bias amplitude and the exact RS position varied although the tip voltage is fixed.[70] By statistically analyzing the RS probability through several times of switching between HRS and LRS by tip scanning, five kinds of RS behaviors are distinguished, as shown in Fig. 13 (e). That means the RS itself is also inhomogeneous even in the grain region of the films.

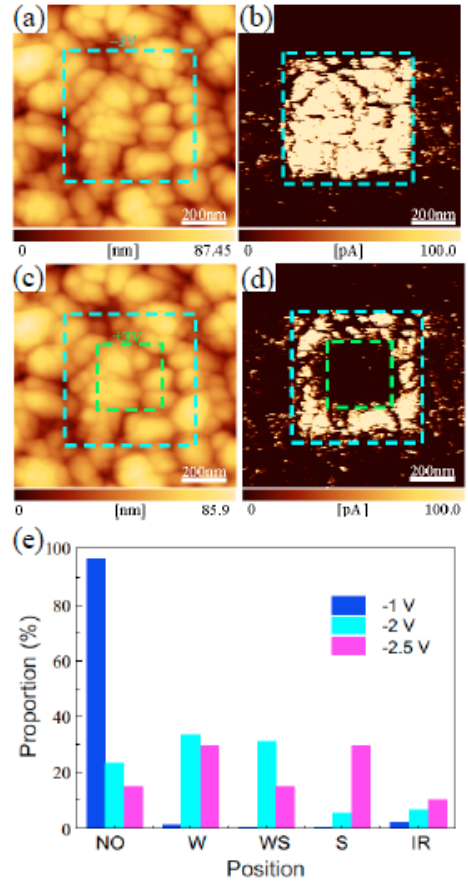

Fig. 13 (a) and (c) show the topographies of the $WO_{3-x}$. (b) and (d) show the corresponding current images after sequential overwriting processes with tip bias of -0.5 V. (e) Proportion of the regions showing different RS behavior. NO: non-RS; W: weak RS; WS: weak to stable RS; S: stable RS; IR: irregular RS.[69,70] Copyright 2011 IOP Publishing Ltd.

The mechanism for the differences in conductivity and RS behaviors between the grain and grain boundary regions is still unclear now. A possible reason might be due to the different oxygen vacancy concentration. This difference could be caused by the more preferential combination of oxygen with oxygen vacancies along the grain boundaries during oxygen annealing process. An oxygen-rich region is expected to form in the grain boundary region, especially on the film surface.[71] Consequently, the conductivity in the grain boundary region is lower than that in the grain region. This difference might cause the non-switching of the grain boundary regions or switching under higher external voltage conditions.

It should be noted that the electric properties of the grain boundary should be different for different oxides. For example, Son *et al* found conductive filaments appeared homogeneously on both the grain and grain boundary region of the NiO films when switching to LRS.[72] After switching to HRS, the filaments in the grain region disappeared, while they still stayed in the grain boundary region. This implies that the grain boundary regions in the NiO films hold a high conductivity but cannot confine the RS behavior in this region. Another example reported by Lanza *et al* discusses polycrystalline $HfO_2$ films, in which RS is only observed at grain boundaries, which exhibit a the low breakdown voltage due to the intrinsic high density of the oxygen vacancies.[73] From the point of view of future memory devices, grain boundary effects on RS will become more technologically important as the dimensions of individual cells in devices continue to shrink and approach the size of individual grains in polycrystalline films. Further investigations to elucidate the grain boundary effect on RS is still required.

**3.2 Electroforming-induced inhomogeneous conductivity in oxides**

The word electroforming comes from an electrochemical process for metallic structure formation, in which metallic elements are plated on the surface of an object at the cathode through redox reactions. In RS investigations, almost all the reported oxides need a pretreatment process before they show the desired RS behavior. This activating process is called electroforming. Generally, electroforming needs an external voltage or current higher than that for the following RS, which causes a reduction of the device resistance. An example is shown in Fig. 14 for ITO/Mg:ZnO/FTO devices.[74] Obviously, three resistance states, the initial resistance state (IRS), the HRS, and the LRS, can be distinguished. RS takes place alternately between the HRS and the LRS, while the IRS cannot be restored after electroforming. By comparing the resistance values on the electrode-area, it is clear that the resistance in the IRS is proportional to the electrode-area. This indicates that the current flows homogenously through the entire electrode area. However, the resistance in both the HRS and LRS is almost independent of the electrode area. This can be explained by the formation of conductive channels, which have a much smaller size than that of the used electrodes. The formation of the conductive channel also causes a redistribution of the oxygen in the film leading to an inhomogeneous distribution after electroforming.

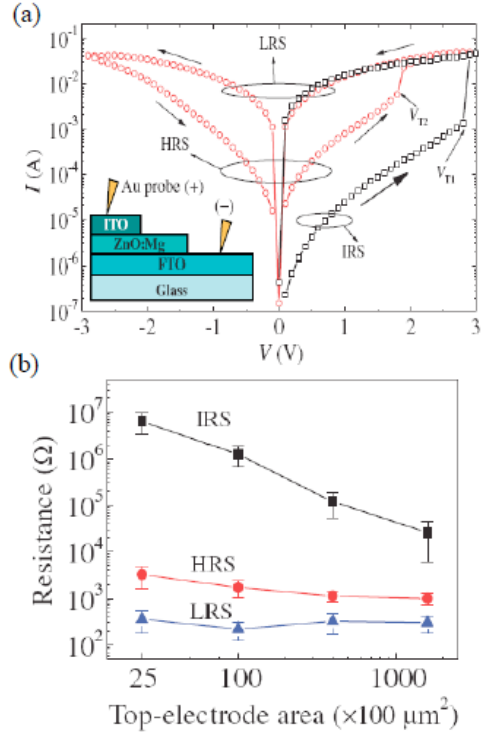

Fig. 14 (a) Semilogarithmic *I-V* characteristics of the indium-tin-oxide (ITO)/Mg:ZnO/F-doped tin oxide (FTO) cell. Inset: Schematic of the measurement. (b) Dependence of resistance on the size of the ITO electrode.[74] Copyright 2009 The Japan Society of Applied Physics.

Despite the importance of electroforming for RS, the details of the process are still unclear. This can be attributed to the many distinct events, which include ionic migration, redox reaction, phase transition, and Joule heating. Their interplay leads to a very complicated overall process. According to the charge type of the active ions in RS, the electroforming can be discussed from two aspects. One is related to the cation migration, which is defined as electrochemical metallization memory (ECM), also called conductive bridging random access memory (CBRAM). Usually the ECM cells consists of an active electrode, such as Ag or Cu, a solid electrolyte, such as chalcogenides or $SiO_2$, and an inert counter electrode such as Pt. In the initial state, ECM cells have a homogeneous conduction and a high resistance values. Waser *et al* have presented an in-depth description of the electroforming process of this type of memory, as shown in Fig. 15(a).[75] During the electroforming process, anodic dissolution of the active electrode takes place to form metal ions, and the metal ions drift across the solid electrolyte towards the inert cathode under the influence of the electric field. Then the metal ions are gradually reduced to metal atoms on the surface of the inert cathode gradually, and accumulate together to form a metallic channel in the solid electrolyte. After electroforming, RS can be realized by the redox process of the metal ions near the active electrode causing the disconnection and connection of the channel. Interesting enough, Yang *et al* found a different electroforming process in the ECM cells recently, where amorphous Si, $SiO_2$ and $Al_2O_3$ thin films were selected as the solid-electrolyte.[76] By both *ex situ* and *in situ* TEM measurements, it was found that the metallic channels growth from the active Ag electrode to the inert Pt electrode (see Fig. 16(g)), being opposite to the direction hypothesized by Waser *et al*. In this case, the following RS is dominated by the region near the inert electrode/electrolyte interface. A similar behavior was also observed in Ag(orCu)/$ZrO_2$/Pt cells.[77] The opposite electroforming behavior has been ascribed to the lower ion mobility in the amorphous Si and oxide electrolytes, leading to an inert reduction of metal ions.

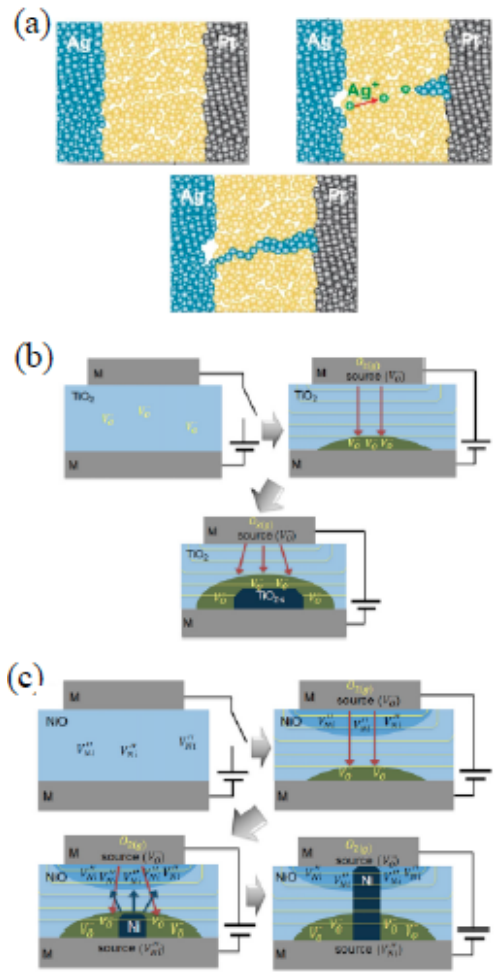

Fig. 15 Schematics of electroforming process in (a) Ag/solid-electrolyte/Pt,[75] Copyright 2011 IOP Publishing Ltd. (b) Pt/$TiO_2$/Pt, c) Pt/NiO/Pt cells.[78] Copyright 2012 IOP Publishing Ltd.

The other aspect of electroforming is related to the migration of the oxygen anion (or oxygen vacancy), which shows a more complicated process than that in the cation case, since the oxygen vacancy migration will induce not only the redox reaction of oxygen itself, but also the gas generation, valence change of metal ions, and even phase transition, and also needs thermal assistance by Joule heating effect. The basic process is the electric-field-induced drift of oxygen vacancies, which gives rise to a virtual cathode made from oxygen rich regions emerging at the real cathode, meanwhile oxygen is released from anode as gas.[45,63] Jeong *et al* reviewed the electroforming processes in binary oxides and give a fully description by separating the binary oxides into two categories: hyper-stoichiometric oxide, such as NiO, and hypo-stoichiometric oxide, such as $TiO_2$, as shown in Fig. 15(b) and (c).[78] In the $TiO_2$ systems, only oxygen vacancy drift is concerned. The anode serves as an oxygen vacancy source. The oxygen vacancies pile up at the cathode, and an oxygen deficient phase, which has been demonstrated to be Magnéli phase,[36,37] grows from the cathode to the anode. In the NiO system, there are two types of defects, oxygen vacancy and nickel vacancy, which coexist in the electroforming. Oxygen and nickel vacancies drift towards the cathode and the anode, respectively, under a voltage. Since no nickel vacancy source exists in the system, the concentration of nickel vacancies near the cathode soon decreases as they migrate towards the anode. Simultaneously, oxygen vacancies from the anode are piled up at the cathode leading to a phase separation into Ni and NiO due to the instability of oxygen-deficient NiO. Then a metallic Ni phase grows from the cathode to the anode until it touches the anode.

Electroforming could even be more complex than presented above, and systematic studies are rather sparse. However, it is clear by now that electroforming introduces a conductivity inhomogeneity into the oxides or makes the oxide conductivity more inhomogeneous than before. Subsequently, reversible RS can be realized. This indicates that the inhomogeneous conductivity inside the oxides should be a prerequisite for RS. In this case, RS behavior, for example, the reported conductive channel connection and disconnection processes, can be considered as a change of oxide conductivity between different inhomogeneous states.

### 3.3 Observation of the conductive channel

The conductive channel describes a special region in the materials, which has a higher conductivity but smaller volume than other regions and carries the most current through the material under an electric field. From the point of view of conductivity distribution, conductive channels can be considered as a typical case of an inhomogeneous conductivity in oxides. Many RS phenomena can be well described by the conductive channel connection/disconnection process. Therefore, the information about the position, dimension, number, composition, structure, and dynamic evolution of the conductive channels during the RS process is a key to understand the physical mechanism as well as a way to improve the switching property.

By now many methods have been used for the conductive channel observation, as summarized in Fig. 16. An early observation of a conductive channel was reported by Hirose et al,[79] who found the dendritic Ag filament in Ag/As$_2$S$_3$/Au cell by a direct optical method. Then a similar result was observed in Ag/AgGeSe/Ni cell by Kozicki et al,[80] and in Ag/H$_2$O/Pt cell by Guo et al.[81] By using the CAFM methods, Szot et al found the channel along the dislocations in a SrTiO$_3$ single crystal and attributed it to the oxygen vacancy migration along the dislocations.[45] Janousch et al investigated the Pt/Cr:SrTiO$_3$ (single crystal)/Pt cell by infrared thermal microscopy, micro X-ray fluorescence (XRF) and x-ray absorption near-edge spectroscopy (XANES) and detected the dumbbell-like channel shape caused by the oxygen vacancy distribution.[82] Yasuhara et al studied a planar-type Pt/CuO/Pt cell by photoemission electron microscopy (PEEM) and found that the channel formed due to the change of the chemical state of Cu ions.[83] Kwon et al using high-resolution TEM combined with CAFM attributed the cone shape channel in TiO$_2$ to the formation of a Magnéli phase.[37] Yang et al also performed TEM measurement on Ag/SiO$_2$/Pt cells and found the Ag filaments growing from the Pt electrode.[76] Recently, the *in situ* observation of conductive channels was realized by using TEM method in Cu/Cu:GeTe/Pt cell, which demonstrated Cu multi-filaments performing during the RS.[84] Since the channel usually has a small size and occurs randomly in oxides, *in situ* monitoring of the dynamic evolution of the whole channel in the progress of RS is difficult but necessary, and will give crucial insights into understanding the detailed channel connection/disconnection process.

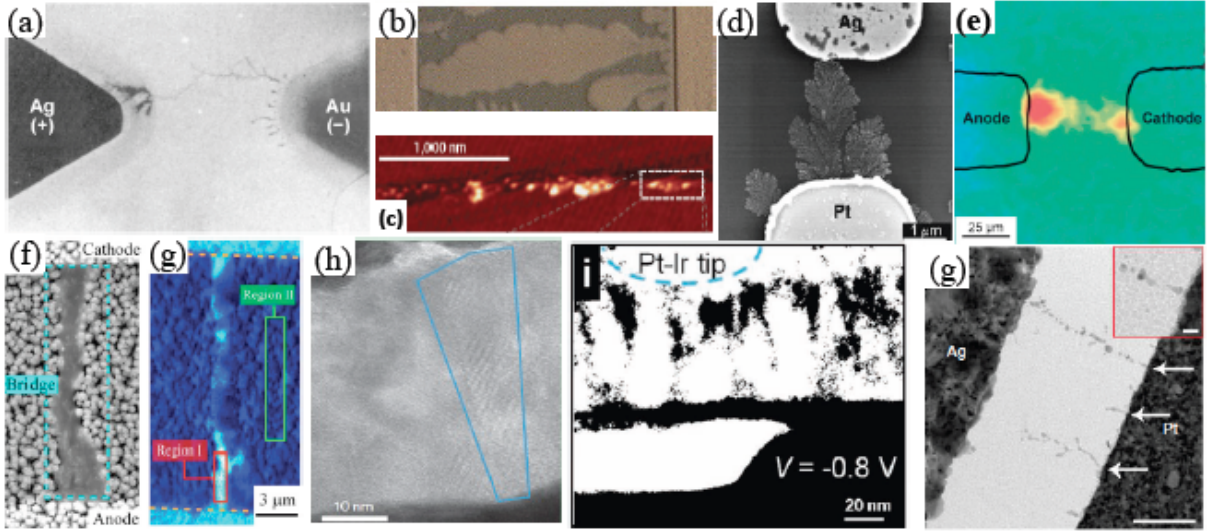

Fig. 16 (a) Conductive channel observation in various materials. (a) Optical microscopy image of a Ag dendrite grown from the Au electrode towards to Ag electrode within a $As_2S_3$ thin film on a glass substrate.[79] Copyright 1976 American Institute of Physics. (b) Optical microscopy image of a Ag filament in Ag/AgGeSe/Ni cell.[80] Copyright 2006 Elsevier. (c) CAFM current mapping of a linear shaped filament in single crystal $SrTiO_3$.[45] Copyright 2006 Nature Publishing Group. (d) A precipitated dendrite Ag filament in $Ag/H_2O/Pt$ cell.[81] Copyright 2007, American Institute of Physics. (e) Cr XRF map taken at 6004.3 eV on the top of a Cr-doped $SrTiO_3$ single crystal with red color scale showing the oxygen vacancy in the Cr octahedral position of the perovskite cell.[82] Copyright 2007 Wiley-VCH. (f) and (g) show SEM image of the planar-type Pt/CuO/Pt cell after the electroforming and corresponding chemical PEEM image, which is obtained from dividing the intensity of PEEM image recorded at photon energies of 932.6 eV by that at 930.3 eV. The photon energies of 930.3 and 932.6 eV correspond to the Cu $L_3$ absorption edges of CuO states and their reduced states ($Cu_2O$ and/or Cu metal).[83] Copyright 2009 American Institute of Physics. (h) High-resolution TEM image showing a cone-shaped conductive channel of a Magnéli ($Ti_4O_7$) phase in a $TiO_2$ matrix.[37] Copyright 2010 Nature Publishing Group. (i) In situ STEM measurement of Cu channel (dark region in the film) in a Pt/Cu-GeTe/Cu cell.[84] Copyright 2011 Wiley-VCH. (g) TEM image of the Ag filaments in a $Ag/SiO_2/Pt$ planar cell.[76] Copyright 2012 Nature Publishing Group.

Besides *in situ* TEM methods, optical imaging method should be an other effective way for the dynamic observation of the channel, since it can conveniently realize the *in situ* measurement as well as the detection of the whole channel region.[85] Fig. 17 shows the conductive channel in a planar $Au/WO_{3-x}/Au$ structure observed by direct optical imaging method.[86] Upon applying a low voltage, the *I-V* curve shows a linear feature, indicative for a good Ohmic-like contact at the $Au/WO_{3-x}$ interface. An obvious hysteresis appeared with increasing voltage, accompanied by a dark region with parabolic shape protruding from the cathode towards the anode. The dark region is considered as the conductive channel due to its higher conductivity compared with the surroundings. Note that, the channel cannot touch the anode, as is expected to penetrate through the oxide before. By changing the external field polarity, a bar-like channel protruded from the right electrode, which is the cathode in this case, and then was pushed into the left parabolic channel. Correspondingly, the device resistance decreased. Reversing the external field polarity again, the bar-like channel shrank back and the device resistance increased again. Obviously, there are four resistance states, which can be distinguished in the Au/WO$_{3-x}$/Au structure, that is, initial state, electroforming state, LRS, and HRS, according to he channel shape and resistance value. Reversible RS can be realized by changing between the LRS and the HRS.

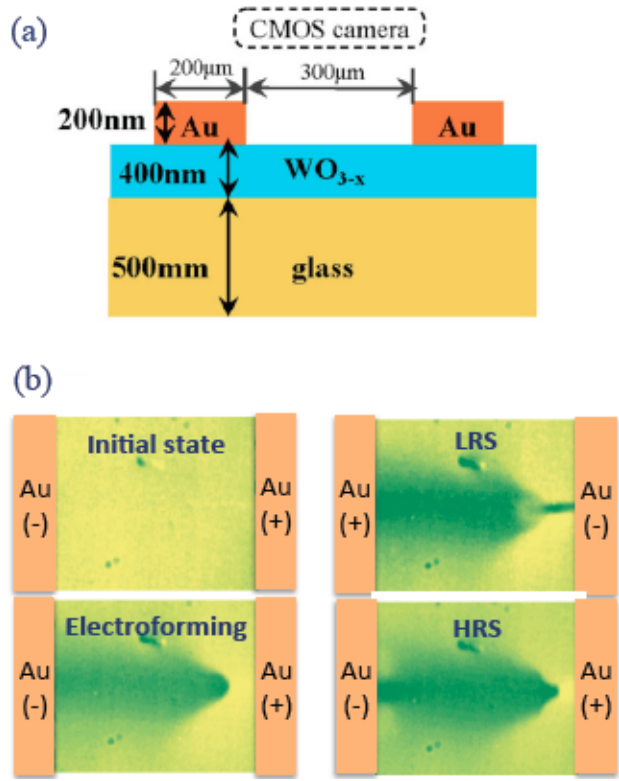

Fig. 17 (a) Schematic of the Au/WO$_{3-x}$/Au planar cell. (b) Images of the conductive channel shapes corresponding to the four resistance states, which appeared in the RS process: initial state, electroforming state, LRS, and HRS.[86] Copyright 2012 American Institute of Physics.

An interesting feature during the RS is that the low-to-high RS is not simply a monotonic process but also dependent on the field application time, as shown in Fig. 18(a). Four typical states were selected to show the conductive channel evolution during the RS. It can be seen that an "are-like" connection between the parabolic- and bar-channels formed the LRS with a relatively low conductive region still existing in the channel center. When performing low-to-high RS, the low conductive region disappeared firstly, but the whole channel was still connected, corresponding to the state C. This geometrical feature explains why the device resistance decreased initially. Then the disconnection of the channel appeared at the right electrode side (i.e. the anode side), corresponding to the state D.

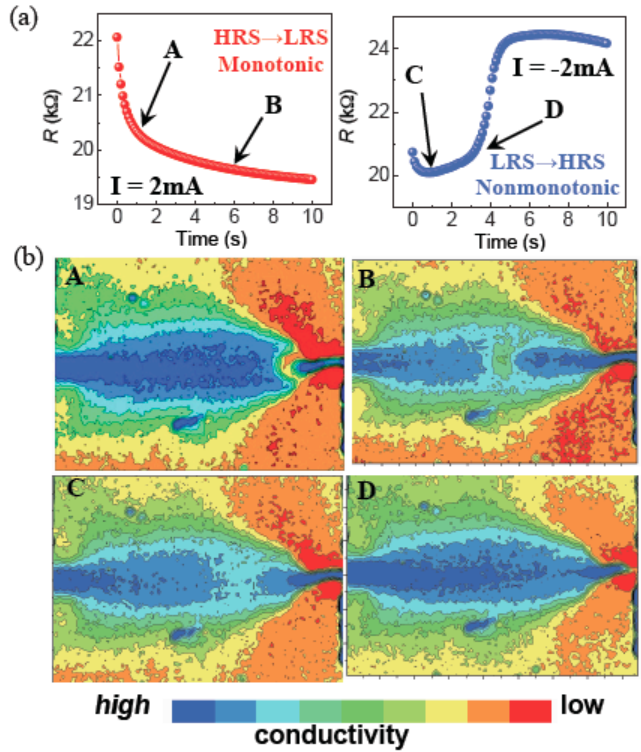

Fig. 18 (a) Time dependent resistance change caused by a constant current with -2 mA (left) and 2 mA (right). (b) Corresponding photo images, snapped simultaneously with resistance measurement at the time of A, B, C, and D, shown in (a).[86] Copyright 2012 American Institute of Physics.

It should be noted that the connection and disconnection between the parabolic channel and bar channel are only required to occur at the region near one of the electrodes, which actually depends on the position of parabolic channel formed during electroforming process. Fig. 19 shows two cells electroformed by electric fields with opposite polarity. It is clear that the RS location, that is, the channel connection/disconnection region, changed from one electrode side to the other. Meanwhile, the RS polarity, if we define one electric filed direction as positive, also reversed. The RS polarity of the planar cell should originate from the asymmetric distribution of the conductivity in the oxide due to the parabolic channel formation in the electroforming process, irrelevant of the interfacial barrier or chemical layer.

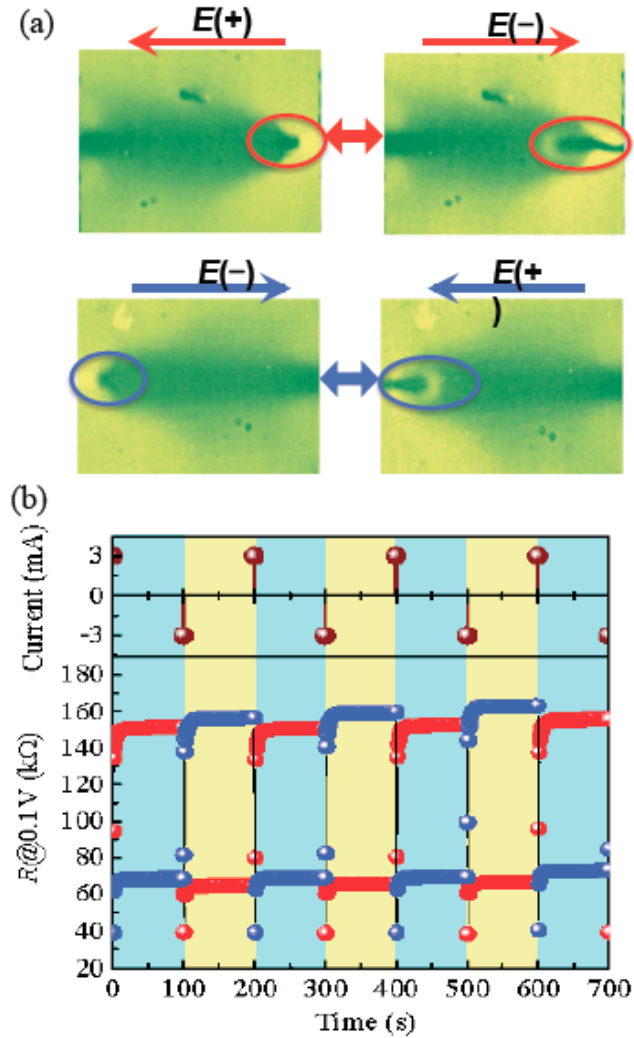

Fig. 19 (a) Photo images of the Au/WO$_{3-x}$/Au planar cell, snapped for the high and low resistance states. The arrows correspond to the polarity direction of the applied electric voltage. The RS regions are marked by several circles. (b) RS of the planar cell triggered by a constant current with an amplitude which alternated between 3 mA and -3 mA for 10 s. Red and blue points correspond to the up-side and down-side images in (a), respectively.

Optical imaging method can also show the conductive channel evolution during instable RS. As shown in Fig. 20, it is obvious that the instable RS, generated by applying a relatively high constant current, corresponds to the occurring position of bar-channel changing gradually from the electrode center to the edge and then coming back. These geometrical features give some insight into the improvement of RS properties. For example, the electric-field distribution in the devices can be concentrated to anchor the bar-like channel formation by the geometrical shape design. Alternatively some active particles can be inserted to induce the conductive channel formation around these particles. It should be noted that the ideas to improve RS performance have been demonstrated in some RS devices.[87-89]

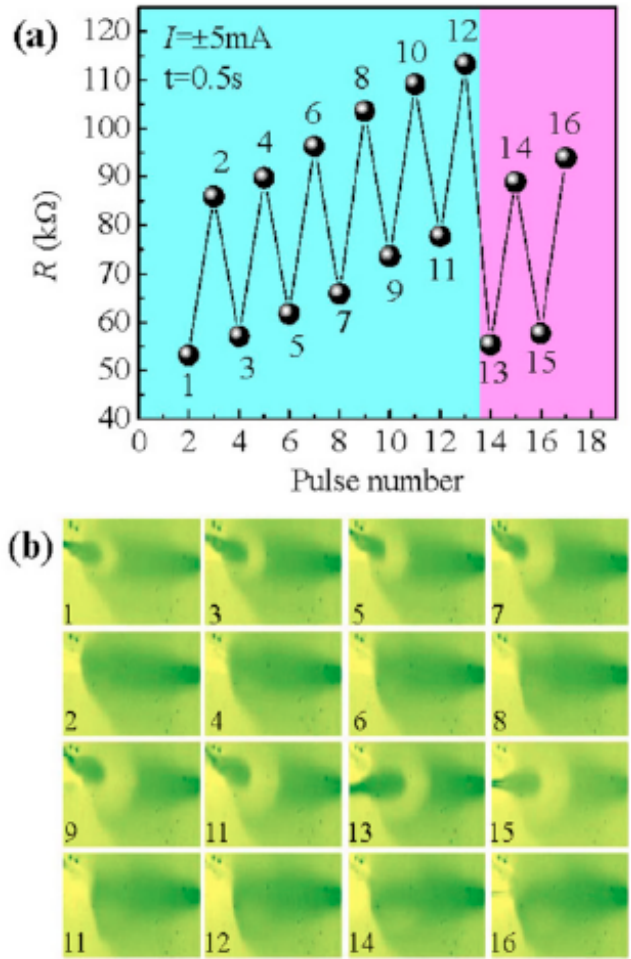

Fig. 20 (a) Instable RS triggered by a constant current of 5 mA with different polarity for 0.5 s. The measuring voltage is 0.1 V. (b) Corresponding photo images snapped simultaneously with resistance measurement.[86] Copyright 2012 American Institute of Physics.

Based on the optical observations, some phenomenological conclusions about the RS process can be given regarding the geometrical features of the conductive channels. 1) The conductive channel is formed in the electroforming process; 2) The conductive channel consists of two types of channels with different shape. The RS is accompanied by the connection/disconnection between the two types of channels. 3) RS only takes place in a region near one of the electrodes determined by the electroforming. 4) The electroforming direction determines the RS polarity. 5) The change of channel location can influence the RS stability. Although the optical imaging method provides a direct observation of the conductive channel in the Au/$WO_{3-x}$/Au cells, the related color change mechanism in is still unclear. No observation of a channel was possible for the same films when a quartz substrate was used. This implies that the color change is due to the sodium migration in the $WO_{3-x}$ films, which changes the valence of tungsten ions. Many aspects, such as the special shape of the channel, Joule heating effect, thermal and electric field distribution, electrode shape influence, and downscaling property, however, still need to be furthered investigated.

## 4. Summary and outlook

This review article summarized the recent RS studies of oxides from the point of view of the conductivity inhomogeneity. For discussion convenience, the RS is divided into two classes, interface RS and bulk RS according to the location of active region. It should be pointed out that this is not a stringent classification and in most cases the RS should be a joint effect of interface and bulk. In the interface RS, the conductive inhomogeneity is present in the interface region. We have discussed the mechanism of the Schottky-like barrier change and interface layer formation due to the chemical reaction between metal electrode and oxides. In the bulk RS, the description of the mechanism has focused on the conductive channel and can be described through the formation, connection and disconnection process of the conductive channel. Although many other RS materials and mechanisms are not included in this review, they are closely related to these two classes. A basic feature for both interface and bulk RS is the conductive inhomogeneity, which can be intrinsic to the oxide itself or introduced by the electroforming process.

Electroforming is an important step to activate the RS behavior by the chemical layer formation at the metal/oxide interface or conductive channel formation in the oxides. Usually the electric field for electroforming should be higher than that for the following RS, and close to the dielectric breakdown field. Thus it is also called "soft-breakdown". Under this condition, an irreversible and extremely non-equilibrium thermalchemical and/or electrochemical reaction will take place, leading to a very stochastic nature of the process, which further results in a large non-uniformity from switching cycle to cycle and from device to device. Therefore, quantitatively manipulating the electroforming or preparing electroforming-free devices provides an opportunity for the improvement of RS performance and uniformity. Taking this consideration into account, some methods for electroforming-free device have been reported, for example, the ultra-thin $HfO_x$ film devices,[90] the ferroelectric-resistive-switching devices with $BiFeO_3$,[91] and the resistive tunnel junction with a thin $AlO_x$ oxide tunnel barrier.[92] As mentioned above, the purpose of electroforming is to create a conductive inhomogeneity in the devices. Therefore, by designing a proper intrinsic conductive inhomogeneity in materials to realize the control of conductive channel formation will be another pathway to solve this problem.

Another important performance indicator is the scalability of RS devices. The local RS behavior appears to be promising to realize the high-density devices, since the conductive channel only needs a very small footprint and leads to a size-independent characteristics. However, the operation current will only reduce slightly upon scaling down the device size, leading to a remarkable increase of the current density required for RS. This generates another challenge for the selector devices, which are needed to constitute the stacked crossbar arrays for increasing memory density.[93] It should be pointed out that the so-called conductive inhomogeneity is defined for a feature size range, in which the operation current will be dependent on the conductivity distribution. For high-density memory devices, therefore, a design of the conductive inhomogeneity on the nanoscale will be required.

As has been said by Ovshinskey, "Information is encoded energy". Each resistance state should correspond to an energetically stable state for RS-based nonvolatile memories, and the switching can be considered as a release or absorption of energy stemming from the electric field, temperature gradient, chemical potential, and so on. Recently, Jeong *et al* classified the nonvolatile memories as either thermodynamic or kinetic memories.[78] The RS behavior is defined to be a consequence of kinetics, that is, one state is in the configuration of the ultimate minimum energy, while the other state should be in a meta-stable configuration with a higher energy. This implies that the conductive inhomogeneity in oxides, e.g. the conductive channels in their host-insulating matrix, is a meta-stable state. One possible reason for this feature might be originated from the inevitable ion motion in the RS mechanism, in which ions act as charged dopants not only moving under the influence of an electric field but also diffusing back under field-free conditions. In this case, the switching between the stable and metal-stable states needs a balance between the switching speed and state stability of the meta-stable state. Hence in future design of RS materials and devices the voltage-time dilemma needs to be overcome.[24]

The RS phenomenon has the potential to be used not only as nonvolatile memory, but also as for embedded reconfigurable logic and artificial synapses to realize hardware-based neuromorphic computing.[94] However, a complete understanding of the RS mechanism is a prerequisite to unravel the potential of this phenomenon. The feature of conductive inhomogeneity discussed in this review only concerns a small fragment of the RS investigations. The RS phenomenon can be classified as a critical phenomenon before the transitional electric breakdown. It occurs under a comparatively high field where many nonlinear effects with corresponding and related chemical, thermal, and electronic phenomena occur. To uncover the veil, further investigations, in particular interdisciplinary studies, are still required.